\documentclass{desyproc}

\begin{document}
\title{Recent Progress with the KWISP Force Sensor}

\author{{\slshape  G. Cantatore$^{1,2}$, A. Gardikiotis$^3$, D.H.H. Hoffmann$^4$, M. Karuza$^{5,2}$, Y.K. Semertzidis$^6$, K. Zioutas$^{3,7}$ }\\[1ex]
$^1$Universit\'a di Trieste, Trieste, Italy\\
$^2$INFN Sez. di Trieste, Trieste, Italy\\
$^3$University of Patras, Patras, Greece\\
$^4$Institut für Kernphysik, TU-Darmstadt, Darmstadt, Germany\\
$^5$Phys. Dept. and CMNST, University of Rijeka, Rijeka, Croatia\\
$^6$Department of Physics, KAIST, Daejeon, Republic of Korea\\
$^7$European Organization for Nuclear Research (CERN), G\'eneve, Switzerland}

\contribID{cantatore\_giovanni}

\confID{11832}  
\desyproc{DESY-PROC-2015-02}
\acronym{Patras 2015} 
\doi  

\maketitle

\begin{abstract}

The KWISP opto-mechanical force sensor has been built and calibrated in the INFN Trieste optics laboratory and is now under off-beam commissioning at CAST. It is designed to detect the pressure exerted by a flux of solar Chameleons on a thin (100 nm) Si$_3$N$_4$ micromembrane thanks to their direct coupling to matter. A thermally-limited force sensitivity of $1.5 \cdot 10^{-14}~\mbox{N}/\sqrt{\mbox{Hz}}$, corresponding to $7.5 \cdot 10^{-16}~\mbox{m}/\sqrt{\mbox{Hz}}$ in terms of displacement, has been obtained. An originally developed prototype chameleon chopper has been used in combination with the KWISP force sensor to conduct preliminary searches for solar chamaleons.

\end{abstract}

\section{Introduction}

The KWISP (Kinetic WISP detection) force-sensor consists of a thin (100 nm) dielectric membrane suspended inside a resonant optical Fabry-Perot cavity \cite{Cantatore1999,Karuza2012,Karuza2013}. The collective force exerted by solar Chameleons bouncing off the membrane surface \cite{Baker2012,Baum2014} will cause a displacement from its equilibrium position which can be sensed by monitoring the cavity resonant frequency. Since, in addition, the membrane is a mechanical resonator, the displacement sensitivity is enhanced by the mechanical quality factor of the membrane. For a detailed description of the KWISP force sensor see \cite{Karuza2015}.
An absolute calibration of the KWISP sensor in terms of force has been obtained in the INFN Trieste optics laboratory by applying a known external force supplied by the radiation pressure of a laser beam ({\em pump beam technique}). This external force is modulated at a given frequency allowing one to explore the frequency region near the mechanical resonance of the membrane. Here we obtain a force sensitivity already at the 300~K thermal limit \cite{Lamoreaux2008}. In order to effectively use the KWISP sensor for chameleon detection it is necessary to find a means of modulating the amplitude of the expected chameleon beam. By exploiting the ability of chameleons to reflect off any material surface when impinging at grazing incidence, and to correspondingly traverse it when at normal incidence \cite{Baker2012}, we have originally devised and built a {\em chameleon chopper} prototype. The chopper allows one to shift the expected chameleon signal away from the noisy region near zero frequency, eventually reaching, with a suitable high frequency chopper, frequencies near resonance. We have used the prototype chopper, working at frequencies below 200 Hz, for preliminary solar chameleon search runs, also taking advantage of the fact that the KWISP membrane orientation in space is such, that a hypothetical chameleon beam from the sun will reflect off it at grazing angles between 0 and 20 degrees for about 1.5 hours each day.
In the following we will briefly describe the sensor setup, the results from absolute calibration measurements, the chameleon chopper prototype and its use in preliminary solar runs.

\section{The KWISP force sensor}
\begin{figure}
\centerline{\includegraphics[width=0.60\textwidth]{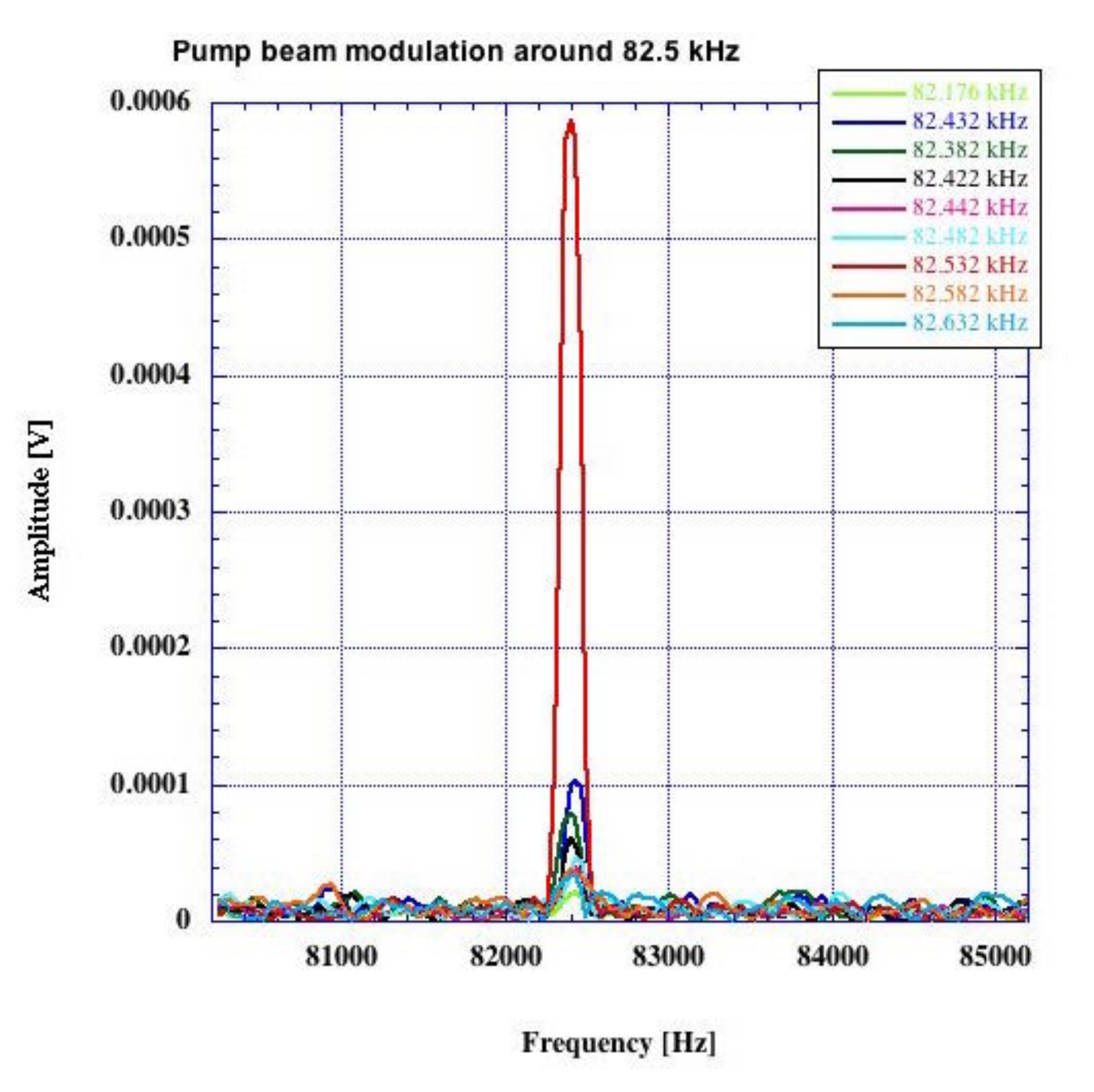}}
\caption{Plot of several power spectra of the error signal from the Fabry-Perot frequency-locking feedback loop. Each spectrum (identified by a uniqe color) has been taken with the pump beam exciting the membrane at a given frequency near the 82.5 kHz membrane resonance frequency (see legend in the figure). Note how the signal amplitude increases when approaching the resonance frequency. From these data one can estimate a mechanical quality factor of $\approx3000$.}\label{Fig:ResFreq}
\end{figure}

The main element of the KWISP force sensor is a vacuum chamber containing an 85 mm long Fabry-Perot cavity made with two 1-inch diameter, 100 cm curvature radius, high-reflectivity, multilayer dielectric mirrors. A  Si$_{3}$N$_{4}$, 5x5 mm$^{2}$, 100 nm thick membrane is inserted inside the cavity and it is initially placed approximately midway between the two cavity mirrors (\emph{membrane-in-the-middle} configuration). The Fabry-Perot cavity is excited using a CW 1064 nm laser beam emitted by a Nd:YAG laser. A second, frequency doubled, CW beam at 532 nm emitted by the same laser is used as an auxiliary beam (\emph{pump beam}) for alignment and for exerting a known external force on the membrane. When the sensor is in detection mode the Fabry-Perot cavity is frequency locked to the laser using an electro-optic feedback loop \cite{Cantatore1999}. The error signal generated by this loop is proportional to the instantaneous frequency difference between laser and cavity and its power spectrum contains the information on membrane displacements.
The pump beam, amplitude-modulated at a given frequency, is then injected into the cavity and it exerts a known force on the membrane by reflecting off it. The intensity of the pump beam corresponds in our case to a net force of $7.9 \cdot 10 ^{-14}$ N.
The presence of this force is detected as a peak in the measured spectrum of the error signal.
The membrane behaves as a mechanical oscillator and its fundamental resonant frequency and quality factor can be directly measured with the pump beam technique. Figure \ref{Fig:ResFreq} shows a plot of several power spectra of the feedback loop error signal. The peaks indicate the presence of the calibration force, while different peaks correspond to different excitation frequencies. The peak with the largest amplitude occurs when the the pump beam modulation frequency matches the membrane mechanical resonance freqeuncy. The background level in Figure \ref{Fig:ResFreq} gives a force sensitivity of $1.5 \cdot 10^{-14}~\mbox{N}/\sqrt{\mbox{Hz}}$, corresponding to $7.5 \cdot 10^{-16}~\mbox{m}/\sqrt{\mbox{Hz}}$ in terms of displacement. These values correspond to the thermal limit at 300 K \cite{Lamoreaux2008}.

\section{Preliminary solar runs with the chameleon chopper}
\begin{figure}
\centerline{\includegraphics[width=0.50\textwidth]{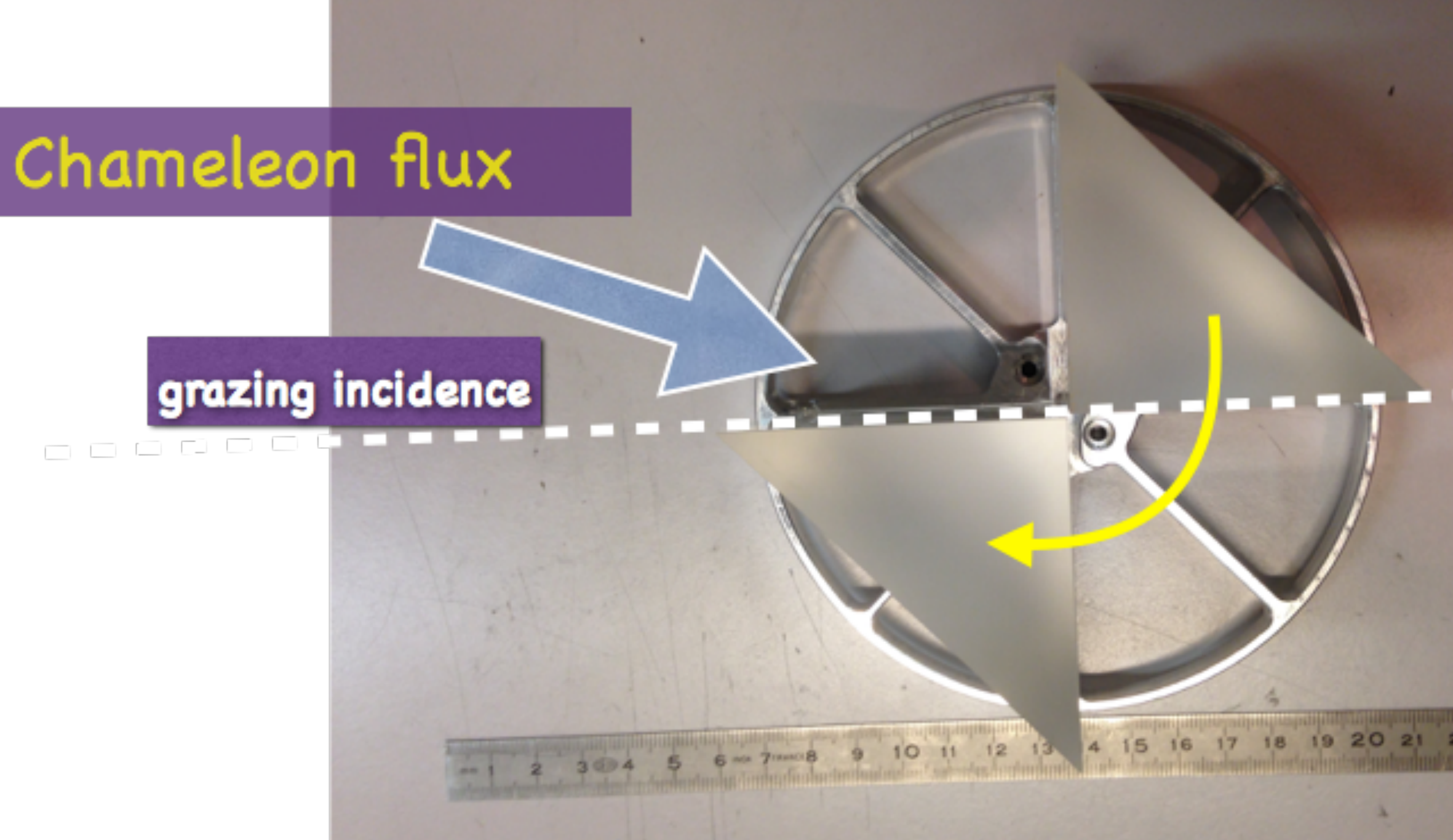}}
\caption{Prototype chameleon chopper and its working principle. The photograph shows a top view of the chameleon chopper prototype consisting of two optical prisms glued to a holding tray capable of rotating along its cylindrical symmetry axis. As the chopper rotates, it presents to the chameleon beam grazing-incidence and normal-incidence surfaces alternatively. The latter transmit chameleons, while the former reflect them, causing the required amplitude modulation. The prototype shown here can rotate at up to $\approx50$ Hz, corresponding to a chopping frequency of $\approx200$ Hz, as a grazing incidence surface is presented to the incoming beam 4 times each turn.}\label{Fig:ChamChop}
\end{figure}

To investigate the possible presence of a signal from a beam of chameleons emitted by the sun, it is necessary to impress a time modulation on it. This can be done by exploiting the general chameleon property of traversing any material when impinging on it at right angles, and of reflecting off it when arriving at grazing incidence (see \cite{Baker2012} for details). We have designed and built a prototype chameleon chopper (see Figure \ref{Fig:ChamChop}) exploiting this property.

\begin{figure}
\centerline{\includegraphics[width=0.70\textwidth]{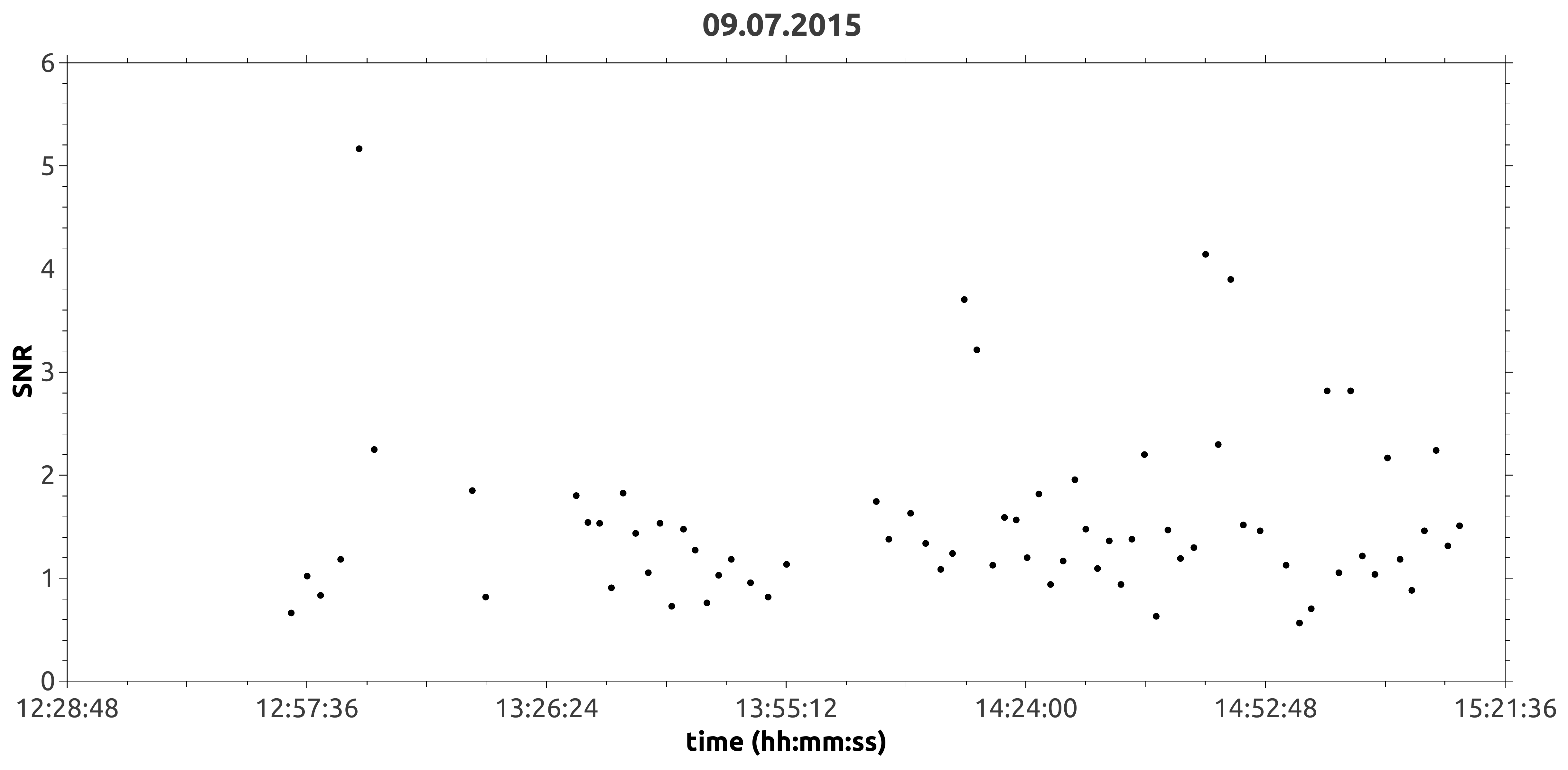}}
\caption{Data from a sample solar run. The graph shows a plot of the Signal-to-Noise ratio (SNR) near the chopper frequency (17 Hz in this case), measured in the power spectrum of the feedback error signal, as a function of time. Data were taken while the sun scanned through grazing incidence angles between 0 and 20 degrees. Notice that the dispersion of the data points indicates the absence of a clear signal.}\label{Fig:SNR}
\end{figure}

The chopper was placed in the proper position in order to intercept a hypothetical solar chameleon beam hitting the membrane at grazing incidence angles between 0 and 20 degrees, depending on the time of day. Data were then acquired by recording 40 s long power spectra of the feedback loop error signal.
A partial preliminary analysis of the solar data was conducted by computing for each spectrum the Signal-to-Noise Ratio and by plotting the SNR as a function of time. A sample plot of this type is shown in Figure \ref{Fig:SNR}.

\section{Conclusions}
The KWISP force sensor now running in the INFN Trieste optics laboratory has been calibrated in absolute terms using a known force exerted by an auxiliary pump beam. The measured sensitivity of $1.5 \cdot 10^{-14}~ \mbox{N}/\sqrt{\mbox{Hz}}$, corresponding to $7.5 \cdot 10^{-16}~\mbox{m}/\sqrt{\mbox{Hz}}$ in terms of displacement, is already at the 300 K thermal limit.
The \emph{chameleon chopper} concept has been implemented in a working prototype \cite{PrepChopper}. This was used in combination with the force-sensor to conduct preliminary runs fro the detection of an hypothetical chameleon beam emitted from the sun. Analysis of the data from these runs is in progress \cite{PrepSolar}. The KWISP force sensor, once coupled to the X-Ray Telescope at CAST, has the potential to access unexplored regions in the Chameleon parameter space, possibly allowing a first glimpse at the nature of Dark Energy \cite{Baum2014}.


\begin{footnotesize}

\end{footnotesize}



\begin{thebibliography}{99}
%

\bibitem{Cantatore1999} G. Cantatore, F. Della Valle, E. Milotti, P. Pace, E. Zavattini, E. Polacco, F. Perrone, C. Rizzo, G. Zavattini, G. Ruoso, Rev. of Sc. Instr. {\bf66(4)}, 2785–2787 (1999).
\bibitem{Karuza2012} M. Karuza, C. Molinelli, M. Galassi, C. Biancofiore, R. Natali, P. Tombesi, G. Di Giuseppe, D. Vitali, New J. of Phys., {\bf14(9)} (2012).
\bibitem{Karuza2013} M. Karuza, M. Galassi, C. Biancofiore, C. Molinelli, R. Natali, P. Tombesi, G. Di Giuseppe, D. Vitali, J. of Optics, {\bf15(2)}, 025704 (2013).
\bibitem{Baker2012} O.K. Baker, A. Lindner, Y. K. Semertzidis, A. Upadhye, K. Zioutas, arXiv:1201.0079 (2012).
\bibitem{Baum2014} S. Baum, G. Cantatore, D.H.H. Hoffmann, M. Karuza, Y.K. Semertzidis, A. Upadhye, K. Zioutas, Physics Letters B {\bf739}, 167–173 (2014).
\bibitem{Karuza2015} M. Karuza, G. Cantatore, A. Gardikiotis, D.H.H. Hoffmann, Y.K. Semertzidis, K. Zioutas, arXiv:1509.04499 (2015).
\bibitem{Lamoreaux2008} S. Lamoreaux, arXiv:0808.4000 (2008).
\bibitem{PrepChopper} K. Zioutas, G. Cantatore, M. Karuza, in preparation.
\bibitem{PrepSolar} G. Cantatore, M. Karuza, K. Zioutas, in preparation.




\end{thebibliography}
\end{document}